\documentclass[sigconf]{acmart}
\AtBeginDocument{%
	\providecommand\BibTeX{{%
			\normalfont B\kern-0.5em{\scshape i\kern-0.25em b}\kern-0.8em\TeX}}}
\setcopyright{none}
\copyrightyear{2019}
\acmYear{2019}
\renewcommand\footnotetextcopyrightpermission[1]{} 
\usepackage{xcolor}
\usepackage{ifthen}
\usepackage{graphicx}
\usepackage{mathtools}
\usepackage{url}
\usepackage{array}
\usepackage[savemem]{listings}
\usepackage{xspace}
\usepackage{tikz}
\usepackage{pgfplots}
\usetikzlibrary{spy,calc}
\usepackage[inline]{enumitem}
\usepackage{subcaption}
\usepackage{cleveref}

\def\Snospace~{\S{}}

\newif\ifshowComments
\showCommentstrue

\newcommand{\todo}[2]{%
	\ifshowComments%
		\ifthenelse{\equal{#1}{MA}}{%
			\todoprint{cyan}{#1}{#2}}{}%
	 	\ifthenelse{\equal{#1}{AI}}{%
	 		\todoprint{magenta}{#1}{#2}}{}%
	 	\ifthenelse{\equal{#1}{AP}}{%
	 		\todoprint{orange}{#1}{#2}}{}%
	 	\ifthenelse{\equal{#1}{}}{%
	 		\todoprint{red}{\color{red} T}{#2}}{}%
	 \fi%
}

\newcommand{\todoprint}[3]{%
\colorbox{#1}{TODO}\footnote{\colorbox{#1}{#2}: #3}%
}
\ifdefined\anonymous
\newcommand{\anonymize}[1]{\underline{\emph{anonymized content}}}
\else
\newcommand{\anonymize}[1]{#1}
\fi

\newif\ifblackandwhitecycle
\gdef\patternnumber{0}
\pgfkeys{/tikz/.cd,
	zoombox paths/.style={
		draw=orange,
		very thick
	},
	black and white/.is choice,
	black and white/.default=static,
	black and white/static/.style={ 
		draw=white,   
		zoombox paths/.append style={
			draw=white,
			postaction={
				draw=black,
				loosely dashed
			}
		}
	},
	black and white/static/.code={
		\gdef\patternnumber{1}
	},
	black and white/cycle/.code={
		\blackandwhitecycletrue
		\gdef\patternnumber{1}
	},
	black and white pattern/.is choice,
	black and white pattern/0/.style={},
	black and white pattern/1/.style={    
		draw=white,
		postaction={
			draw=black,
			dash pattern=on 2pt off 2pt
		}
	},
	black and white pattern/2/.style={    
		draw=white,
		postaction={
			draw=black,
			dash pattern=on 4pt off 4pt
		}
	},
	black and white pattern/3/.style={    
		draw=white,
		postaction={
			draw=black,
			dash pattern=on 4pt off 4pt on 1pt off 4pt
		}
	},
	black and white pattern/4/.style={    
		draw=white,
		postaction={
			draw=black,
			dash pattern=on 4pt off 2pt on 2 pt off 2pt on 2 pt off 2pt
		}
	},
	zoomboxarray inner gap/.initial=5pt,
	zoomboxarray columns/.initial=2,
	zoomboxarray rows/.initial=2,
	subfigurename/.initial={},
	figurename/.initial={zoombox},
	zoomboxarray/.style={
		execute at begin picture={
			\begin{scope}[
				spy using outlines={%
					zoombox paths,
					width=\imagewidth / \pgfkeysvalueof{/tikz/zoomboxarray columns} - (\pgfkeysvalueof{/tikz/zoomboxarray columns} - 1) / \pgfkeysvalueof{/tikz/zoomboxarray columns} * \pgfkeysvalueof{/tikz/zoomboxarray inner gap} -\pgflinewidth,
					height=\imageheight / \pgfkeysvalueof{/tikz/zoomboxarray rows} - (\pgfkeysvalueof{/tikz/zoomboxarray rows} - 1) / \pgfkeysvalueof{/tikz/zoomboxarray rows} * \pgfkeysvalueof{/tikz/zoomboxarray inner gap}-\pgflinewidth,
					magnification=3,
					every spy on node/.style={
						zoombox paths
					},
					every spy in node/.style={
						zoombox paths
					}
				}
				]
			},
			execute at end picture={
			\end{scope}
			\node at (image.north) [anchor=north,inner sep=0pt] {\subcaptionbox{\label{\pgfkeysvalueof{/tikz/figurename}-image}}{\phantomimage}};
			\node at (zoomboxes container.north) [anchor=north,inner sep=0pt] {\subcaptionbox{\label{\pgfkeysvalueof{/tikz/figurename}-zoom}}{\phantomimage}};
			\gdef\patternnumber{0}
		},
		spymargin/.initial=0.5em,
		zoomboxes xshift/.initial=1,
		zoomboxes right/.code=\pgfkeys{/tikz/zoomboxes xshift=1},
		zoomboxes left/.code=\pgfkeys{/tikz/zoomboxes xshift=-1},
		zoomboxes yshift/.initial=0,
		zoomboxes above/.code={
			\pgfkeys{/tikz/zoomboxes yshift=1},
			\pgfkeys{/tikz/zoomboxes xshift=0}
		},
		zoomboxes below/.code={
			\pgfkeys{/tikz/zoomboxes yshift=-1},
			\pgfkeys{/tikz/zoomboxes xshift=0}
		},
		caption margin/.initial=4ex,
	},
	adjust caption spacing/.code={},
	image container/.style={
		inner sep=0pt,
		at=(image.north),
		anchor=north,
		adjust caption spacing
	},
	zoomboxes container/.style={
		inner sep=0pt,
		at=(image.north),
		anchor=north,
		name=zoomboxes container,
		xshift=\pgfkeysvalueof{/tikz/zoomboxes xshift}*(\imagewidth+\pgfkeysvalueof{/tikz/spymargin}),
		yshift=\pgfkeysvalueof{/tikz/zoomboxes yshift}*(\imageheight+\pgfkeysvalueof{/tikz/spymargin}+\pgfkeysvalueof{/tikz/caption margin}),
		adjust caption spacing
	},
	calculate dimensions/.code={
		\pgfpointdiff{\pgfpointanchor{image}{south west} }{\pgfpointanchor{image}{north east} }
		\pgfgetlastxy{\imagewidth}{\imageheight}
		\global\let\imagewidth=\imagewidth
		\global\let\imageheight=\imageheight
		\gdef\columncount{1}
		\gdef\rowcount{1}
		
	},
	image node/.style={
		inner sep=0pt,
		name=image,
		anchor=south west,
		append after command={
			[calculate dimensions]
			node [image container,subfigurename=\pgfkeysvalueof{/tikz/figurename}-image] {\phantomimage}
			node [zoomboxes container,subfigurename=\pgfkeysvalueof{/tikz/figurename}-zoom] {\phantomimage}
		}
	},
	color code/.style={
		zoombox paths/.append style={draw=#1}
	},
	connect zoomboxes/.style={
		spy connection path={\draw[draw=none,zoombox paths] (tikzspyonnode) -- (tikzspyinnode);}
	},
	help grid code/.code={
		\begin{scope}[
			x={(image.south east)},
			y={(image.north west)},
			font=\footnotesize,
			help lines,
			overlay
			]
			\foreach \x in {0,1,...,9} { 
				\draw(\x/10,0) -- (\x/10,1);
				\node [anchor=north] at (\x/10,0) {0.\x};
			}
			\foreach \y in {0,1,...,9} {
				\draw(0,\y/10) -- (1,\y/10);                        \node [anchor=east] at (0,\y/10) {0.\y};
			}
		\end{scope}    
	},
	help grid/.style={
		append after command={
			[help grid code]
		}
	},
}

\newcommand\phantomimage{%
	\phantom{%
		\rule{\imagewidth}{\imageheight}%
	}%
}
\newcommand\zoombox[2][]{
	\begin{scope}[zoombox paths]
		\pgfmathsetmacro\xpos{
			(\columncount-1)*(\imagewidth / \pgfkeysvalueof{/tikz/zoomboxarray columns} + \pgfkeysvalueof{/tikz/zoomboxarray inner gap} / \pgfkeysvalueof{/tikz/zoomboxarray columns} ) + \pgflinewidth
		}
		\pgfmathsetmacro\ypos{
			(\rowcount-1)*( \imageheight / \pgfkeysvalueof{/tikz/zoomboxarray rows} + \pgfkeysvalueof{/tikz/zoomboxarray inner gap} / \pgfkeysvalueof{/tikz/zoomboxarray rows} ) + 0.5*\pgflinewidth
		}
		\edef\dospy{\noexpand\spy [
			#1,
			zoombox paths/.append style={
				black and white pattern=\patternnumber
			},
			every spy on node/.append style={#1},
			x=\imagewidth,
			y=\imageheight
			] on (#2) in node [anchor=north west] at ($(zoomboxes container.north west)+(\xpos pt,-\ypos pt)$);}
		\dospy
		\pgfmathtruncatemacro\pgfmathresult{ifthenelse(\columncount==\pgfkeysvalueof{/tikz/zoomboxarray columns},\rowcount+1,\rowcount)}
		\global\let\rowcount=\pgfmathresult
		\pgfmathtruncatemacro\pgfmathresult{ifthenelse(\columncount==\pgfkeysvalueof{/tikz/zoomboxarray columns},1,\columncount+1)}
		\global\let\columncount=\pgfmathresult
		\ifblackandwhitecycle
		\pgfmathtruncatemacro{\newpatternnumber}{\patternnumber+1}
		\global\edef\patternnumber{\newpatternnumber}
		\fi
	\end{scope}
}

\begin{document}

\title{Taxonomy-as-a-Service: How To Structure Your Related Work}


\author{Mohsen Ahmadvand, Amjad Ibrahim, and Felix Huber}
	\affiliation{Technische Universit\"at M\"unchen}
		\email{firstname.lastname@cs.tum.edu}

\renewcommand{\shortauthors}{Ahmadvand et al.}
\begin{abstract}
Structuring related work is a daunting task encompassing literature review, classification,
comparison (primarily in the form of concepts), and gap analysis.
Building taxonomies is a compelling way to structure concepts in the literature yielding reusable and extensible models.
However, constructing taxonomies as a product of literature reviews could become, 
to our experiences, immensely complex and error-prone. 
Including new literature 
or addressing errors may cause substantial changes (ripple effects) in taxonomies
coping with which requires adequate tools. 
To this end, we propose a \emph{Taxonomy-as-a-Service (TaaS)} platform. 
TaaS combines the systematic paper review process with 
taxonomy development, visualization, and analysis capabilities. 
We evaluate the effectiveness and efficiency of our platform by employing
it in the development of a real-world taxonomy.
Our results indicate that our TaaS can be used to effectively craft and maintain UML-conforming 
taxonomies and thereby structure related work.
The screencast of our tool demonstration is available at \url{https://goo.gl/GsTjsP}.
 
\end{abstract}
%

\keywords{Taxonomy, Ontology Visualization, Research Tools}

\maketitle
\section{Introduction}
\label{sec:introduction}
Researchers often produce a taxonomy (ontology)\footnote{Although ontologies express more complex relations between concepts than taxonomies, we will use the two terms interchangeably.} that abstracts concepts found in the published literature, around a specific topic, and relate them.  A taxonomy aids its constructor in coping with the growing amount, and complexity of concepts found in the literature, and hence, facilitates a thorough literature review process. 
Taxonomies serve as a communication tool supporting the understandability of concepts.

Researchers usually  model different views of their research domain in a taxonomy. 
We refer to each view as a \textit{taxonomy dimension}. 
A \textit{dimension} groups the concepts related to a specific artifact or a perspective on the research topic. 
For instance, in the security domain, researchers may structure concepts from the \textit{attacker or defender} perspectives \cite{ahmadvand2018taxonomy}, 
or according to  \textit{the what and the how} aspects of protection methods. 
A \textit{dimension} may, also, reflect a process view of the system in which each dimension abstracts a specific phase. 

Primarily, a taxonomy is comprised of a set of interrelated concepts.
There are two types of relationships among concepts - inter-relations (the relationship among concepts in different dimensions) and intra-relations (relations amongst concepts within an arbitrary dimension).
While there exist a wide range of relation types,  
UML relations seem to support a sufficient set of semantics to express a wide range of taxonomies \cite{parreiras2010visualizing}.
Particularly, class diagrams with their built-in relations, viz. association, inheritance, 
composition, and aggregation are good candidates for modeling taxonomy dimensions \cite{ahmadvand2018taxonomy}.
Further refined relations are also possible by  annotating the given relationships.  

Crafting a taxonomy starts with a literature review.
Two systematic review methodologies are widely practiced in the research community:  SLR~\cite{Kitchenham07guidelinesfor} and SMS~\cite{petersen2008systematic}.
They are time-consuming, require substantial  manual effort, and error-prone. 
Hence, automating all (or parts) of them is beneficial. 
Withstanding the differences in the process, 
literature review in essence supplies concepts in the field and their relations upon which a taxonomy is built.
The missing element here is the tool support for crafting taxonomies as the outcome of reviews.

After constructing a  taxonomy, researchers analyze it thoroughly and keep on maintaining and evolving it. These activities are strikingly complex and error-prone as the number of concepts and papers increases. 
Fixing errors such as misclassification, duplicates, and overlooked concepts could render all the previously gathered reports (analyses) obsolete. 




\textbf{Gaps.} 
To the best of our knowledge, the gaps in the literature (see \Cref{sec:related} for the related work) are:
\textbf{i)} there is a lack of adequate tool-support for developing and maintaining taxonomies as a product of SLR or SMS; and
\textbf{ii)} the existing tools rather offer limited structural and gap analysis tools, and they do not facilitate the process of correcting and extending taxonomies.

\textbf{Contributions.}
Our contributions are manifold: \textbf{i)} elicit requirements for a taxonomy development and maintenance service for crafting UML-like taxonomies;
	\textbf{ii)} propose an architecture complying with the elicited requirements;
	\textbf{iii)} develop an interactive visualization tools for  crafting and analyzing taxonomies;
	\textbf{iv)} a thorough evaluation of the tool using a real-world taxonomy; and
	\textbf{v)} open source the entire tool chain.

\section{Requirements}\label{sec:requirements}

In this section, we elicit the system requirements in accordance with the process proposed in~\cite{zowghi2005requirements}.
Space limitation only allows us to list the requirements.  

\subsection{Functional Requirements}\label{subsec:fr}
\textit{FR1.} The system should provide users with a workspace to review papers, create, and update taxonomies. 

\textit{FR2.} Present a mechanism to import the literature to be reviewed. 
Interfaces to upload different formats of literature (e.g., PDF or DOI) should be supported.

\textit{FR3.} Facilitate defining, editing, merging, and relating concepts by multiple researchers. 

\textit{FR4.} Support creating a multidimensional visual model of the identified concepts. 
UML relations of type \textit{association}, \textit{inheritance,} and \textit{composition} shall be supported and distinctly visualized. 
Annotation is also supported to constitute more specialized relations. 

\textit{FR5.} Enable correlating different concepts, and displaying the literature coverage around them.  

\textit{FR6.} Support clutter-free visualizations of the hierarchy of the concepts via 2D and 3D matrix views with zoom and filtering features. 

\textit{FR7.} 
Enable mass literature mapping using keyword matching techniques to update existing taxonomies.

\subsection{Non-Functional Requirements}\label{subsec:nfr}


\textit{NFR1. Scalability, Multi-tenancy, and Deployability.} 
Since the published articles are significantly growing over years~\cite{reller2016morepublishing}, the system should scale up and down based on the load. 

\textit{NFR2. Security.} Secure accesses to unpublished research artifacts.

\textit{NFR3. Fast viewing.} Render taxonomy views by keeping caches of highly demanded visualizations.

\section{Design}\label{sec:design}


\subsection{Architecture}
To completely satisfy \emph{NFR1}, for the architecture of our system, we resort to microservice-based 
architecture. 
It also partially addresses security requirements, \emph{NFR2}, (e.g., isolation and authentication)  as we discuss in \Cref{sec:implementation}.

\begin{figure}
	\centering
	\includegraphics[clip, trim=0.45cm 0.5cm 0.6cm 0.5cm,width=0.35\textwidth]{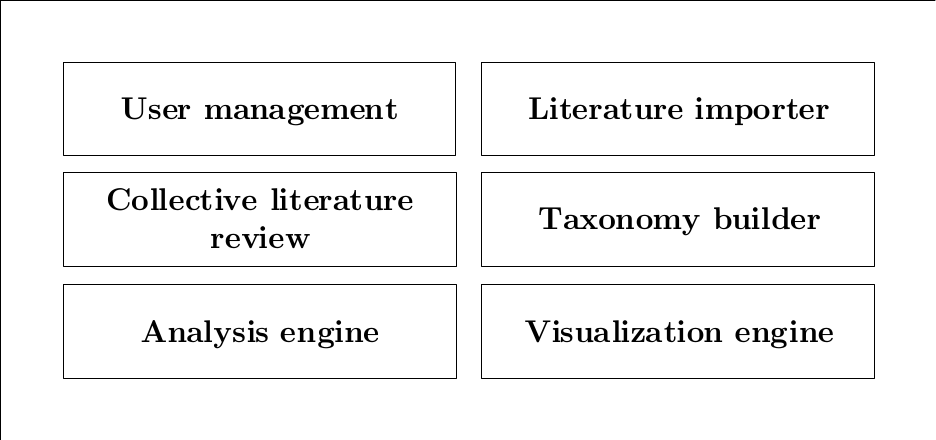}
	\caption{TaaS microservices}
	\label{fig:design-architecture}
	\vspace*{-.5cm}
\end{figure}

As depicted in \Cref{fig:design-architecture}, our microservices are 
- \emph{user management}, \emph{collective literature survey}, \emph{literature importer}, 
\emph{taxonomy builder}, \emph{analysis engine}, and \emph{visualization engine}.

\subsection{Taxonomy Development Process}

The process starts by formulating research questions and keywords. 
It is followed by gathering literature with the specified keywords. 
The collected \emph{articles} are then input into the system. 
As the first step, they are fed into \emph{collective review} microservice whereby 
researchers vote on the relevance of the articles to the research questions of interest.
During the review process papers are marked with a set of classification tags, 
which could be imported as (preliminary) concepts to a taxonomy.
The \emph{taxonomy builder} then enables researchers to extend the preliminary classifications further.

Once a taxonomy is crafted, users can utilize the \emph{analysis and visualization engines} 
for a thorough analysis, or compile reports. 
From this point on, using \emph{literature importer} new papers can be mapped 
to existing (concepts) taxonomies based on the provided keyword matching techniques in the platform.
\begin{figure}[!htb]
	\centering
	\includegraphics[clip, trim=0.45cm 0.5cm 0.6cm 0.5cm,width=0.9\linewidth]{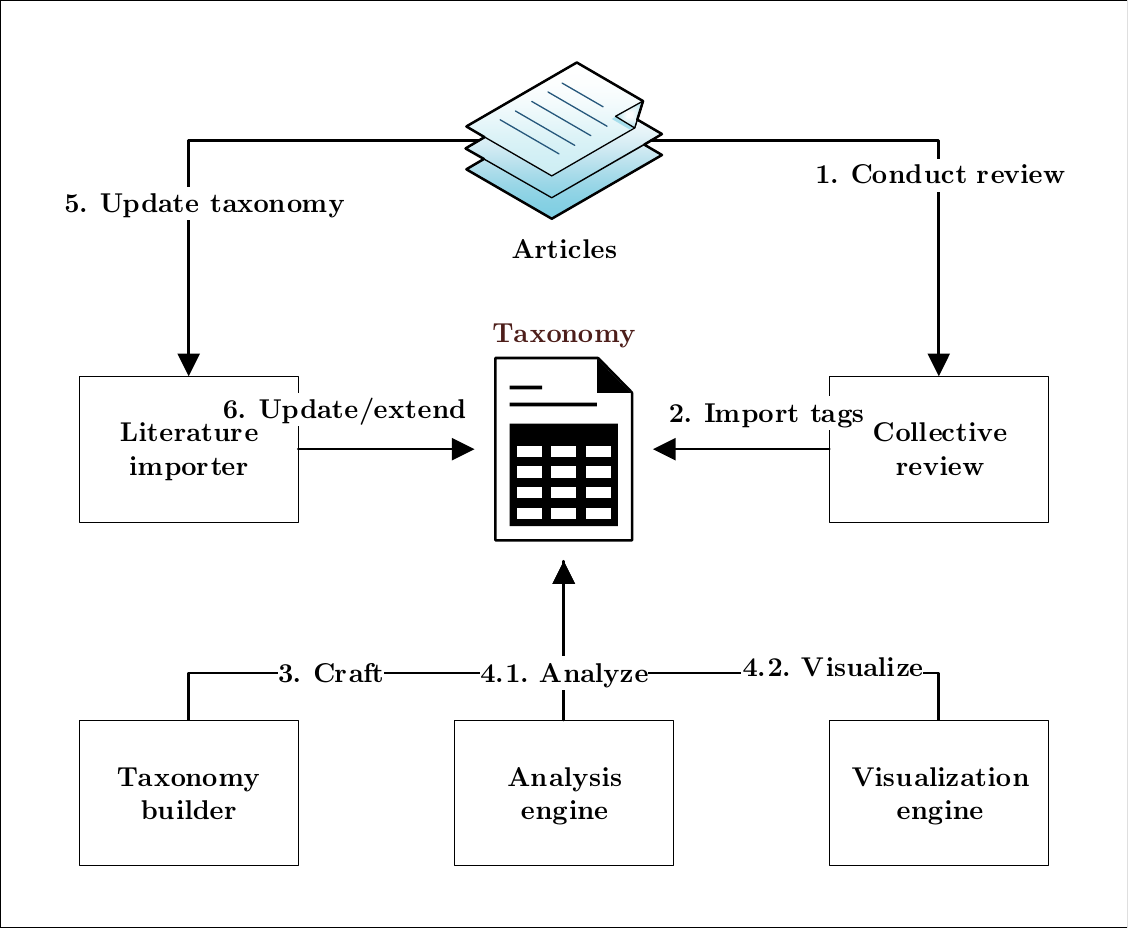}
	\caption{TaaS process}
	\label{fig:process}
\end{figure}
\subsection{Services}
\subsubsection{User management}
Essentially, this service handles user authentications by issuing access tokens 
enabling them to interact with other services in the system.
This service together with other utilized technologies in the implementation of our services
(see \Cref{sec:implementation}) addresses \emph{NFR2}.
\subsubsection{Collective literature survey}
Once researchers gathered the related work (from various sources), they import them into the survey service.
The service then allows coworkers (researchers) to conduct a collective review in which they review the abstract of papers and vote to include or exclude them, 
based on their relevance to the research questions of interest.
The approved papers can be fetched at any time by specifying the minimum number of positive votes. 
Such papers are then analyzed (read) in-depth by individual researchers for final decision makings.
Papers can be tagged with arbitrary keywords as well as notes. 
These keywords could later directly be translated to concepts (in a taxonomy), 
or be used to derive other concepts.
This service in part addresses \emph{FR1, FR2, and FR3} requirements.

\subsubsection{Taxonomy builder}\label{sec:design:subsec:taxonomy-builder}
The builder itself is comprised of three components - \emph{inter-dimensional editor}, 
\emph{intra-dimension editor}, and \emph{tag-to-concept importer}.

\emph{Inter-dimensional editor:} This service enables users to create the dimensions of a taxonomy along with their inter-relationships. 
It is the view in which all the concepts of each dimension and their inter-relations with other dimensions are created and maintained over time.
The inter-dimensional view captures a high-level notion of the taxonomy.
However, each dimension, specific to a particular aspect of the field, needs to be further developed on its own.

\emph{Intra-dimensional editor:}
In a sense, the intra-dimension service provides a zoomed-in view of a dimension of interest, 
whereby all the concepts in a dimension are extended with their (sub) concepts and their further instantiations.
Relationships between concepts can be defined in the form of UML relations 
(aggregation, composition, inheritance, and association).
All the relations support annotations to capture arbitrary semantics. 
Moreover, \emph{fork} and \emph{merge} features are supported to deal 
with the potential mistakes that are caused by the collectively gathered tags, which contribute to addressing \emph{FR3}.
In all the operations of the editor, we utilize an eventual cache consistency policy to honor NFR3. 

\emph{Tag-to-concept importer:} 
Tagged papers throughout the review process can directly be imported into a taxonomy.
This service contributes to addressing \emph{FR1, FR3, and FR4} requirements.
\subsubsection{Literature importer}
Using this service one can upload recent/newly discovered literature to update a taxonomy with the latest literature. 
The service provides four keyword matching methods 
- \emph{regex}, \emph{dice coefficient}, \emph{Levenshtein distance} \cite{gomaa2013survey}, and \emph{fuzzy sort}\footnote{https://github.com/farzher/fuzzysort} for a 
preliminary mapping of papers to the concepts in a taxonomy (\emph{FR7}). 
Researchers can further refine the suggested mappings in the process.
\subsubsection{Analysis engine}
The two core analyses are the \emph{correlation generator} and the \emph{filtering service}.
This service contributes to addressing \emph{FR5 and FR6} requirements.
\subsubsection{Visualization engine}
To aid the development and understandability of taxonomies the visualization engine supports three distinct techniques, 
viz. \emph{hierarchy-matrix}, \emph{3D}, and \emph{Crop-circles}\cite{Wang2006CropCirclesTS} visualizations.
The hierarchy-matrix view combines a matrix visualization with a hierarchical tree view of the taxonomy.
Every cell in the matrix reports the number of papers that are mapped to  both concepts corresponding to x and y-axes. 
The \emph{3D view} extends the matrix view by mapping an arbitrary property (such as the  number of citations) to the $z-$axis.

The \emph{Cropcircles visualization}  offers a clear hierarchy of the concepts 
that have parent-child relationships grouped by their corresponding dimensions.
Therefore, it provides  a better understanding of a taxonomies' topology.
Users can zoom into circles to explore related concepts. 

All the views offer export as images (in PNG format).
The visualization service contributes to addressing \emph{FR6 and FR7} requirements.

\section{Implementation}\label{sec:implementation}
The entire platform (written in Go, MySQL, and HTML5) is made open source and is publicly available on 
Github at \url{https://github.com/mr-ma/paper-review-go}.

\subsection{Modules}
\begin{figure}[!htb]
	\includegraphics[clip, trim=0.45cm 0.5cm 0.6cm 0.5cm,width=\linewidth]{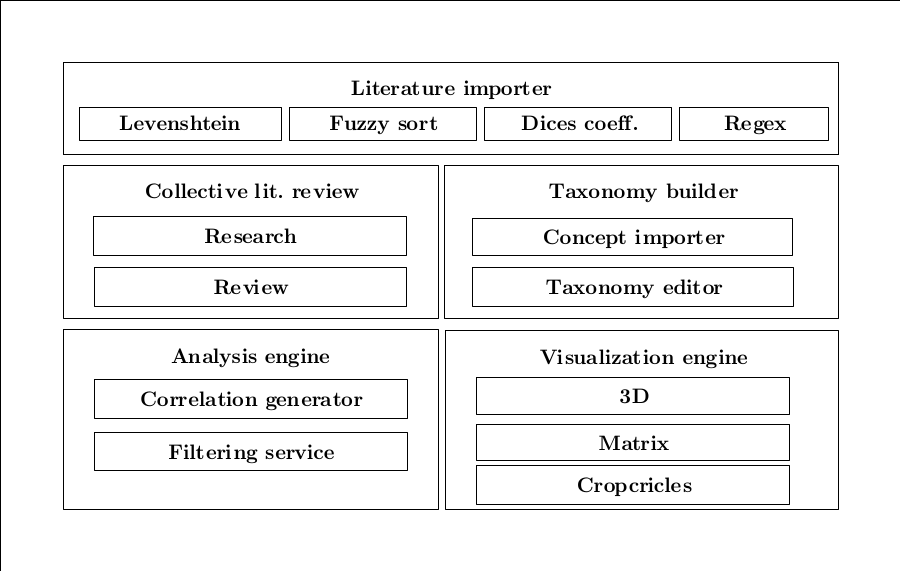}
	\caption{TaaS Modules}
	\label{fig:implementation-modules}
\end{figure}

We split each microservice into a set of goal-oriented modules according to the proposed design (see \Cref{sec:design}). 
\Cref{fig:implementation-modules} captures our modules per microservice. 


\subsection{Deployment}
\begin{figure}[!htb]
	\includegraphics[width=\linewidth]{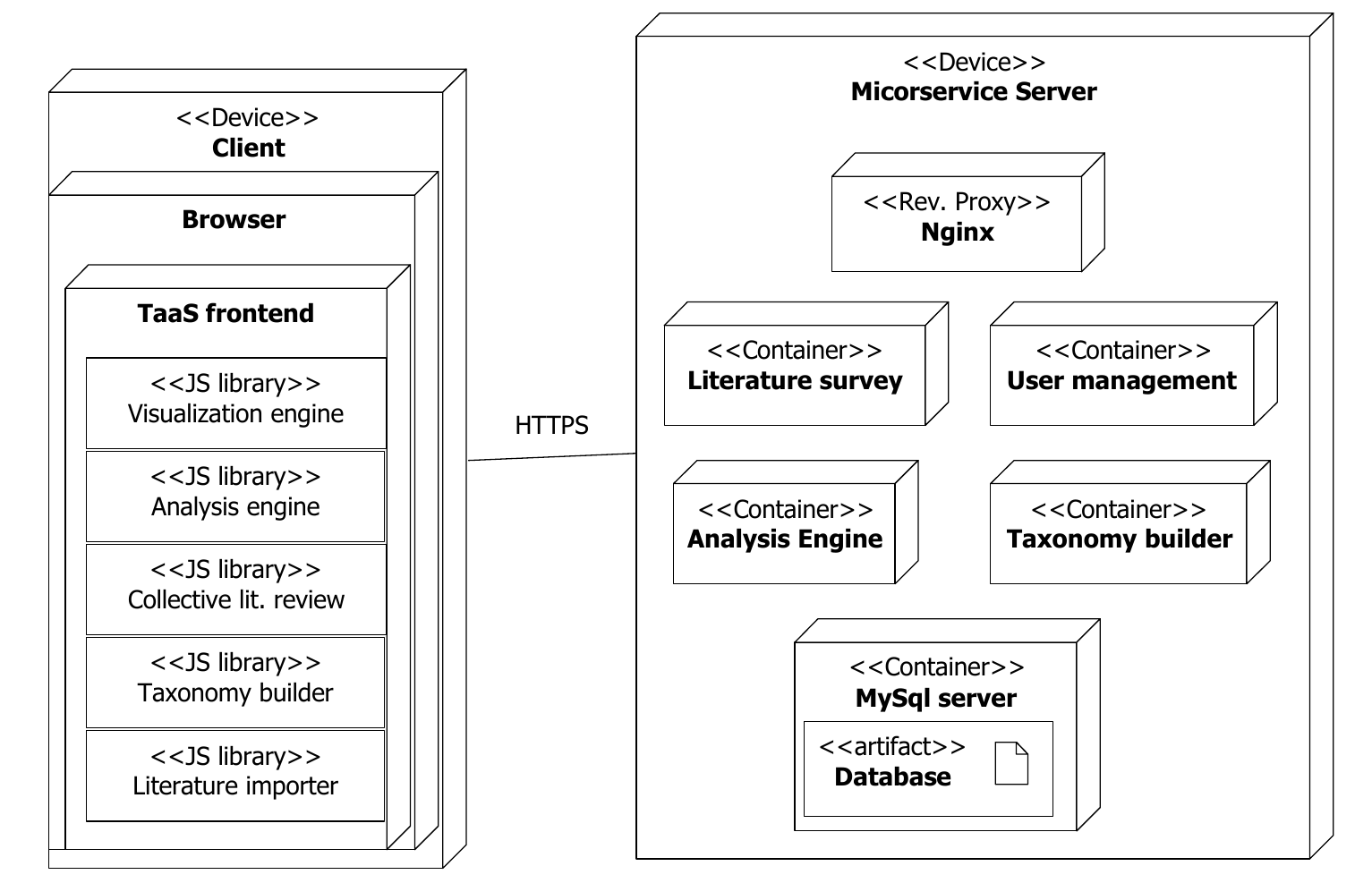}
	\caption{TaaS deployment diagram}
	\label{fig:implementation-deployment}
\end{figure}

For the deployment of our services, we utilize container-based approach (one container per microservice). 
This guarantees a conflict-free service deployment and offers better service isolation.
Scaling the system in this setting is as simple as spinning new containers for microservices under stress (NFR1). \Cref{fig:implementation-deployment} depicts the deployment diagram of our TaaS. 




\section{Evaluation}\label{sec:evaluation}
%

\subsection{Case study: software integrity protection taxonomy}
As an empirical evaluation, we attempt to craft the already existing software integrity protection taxonomy~\cite{ahmadvand2018taxonomy} using our TaaS. 

In their publication, the authors present three different views of their taxonomy, 
viz. a 3-dimensional view with a zoomed-in view of each dimension, 
a matrix view, and eight correlation views.
%
%
We were able to plot the three views and correlations successfully. 
Space limitation hinders enclosing the generated figures as results of these steps.

\subsection{Efficiency}
To carry out performance measurements, 
we use a MacBook Pro machine running macOS High Sierra 10.13 64-bit with Intel i5 2.90 GHz CPU and 16 GB of Ram.

%
We notice that the matrix view incorporates all citations and concepts in a taxonomy
and thus it could potentially underperform as the size of the taxonomy grows. All other views perform linearly. 
%
\subsubsection{Matrix creation}
To identify upper bounds, we measure the elapsed time in the creation of a set of $n\times n$ matrices, where $10<=n<=200$.
We randomly create these matrices initialized with dummy concepts half of which are set to be correlated. 

\begin{figure}[!htb]
	\includegraphics[clip, trim=0.0cm 0.0cm 0.0cm 0.0cm,width=\linewidth]{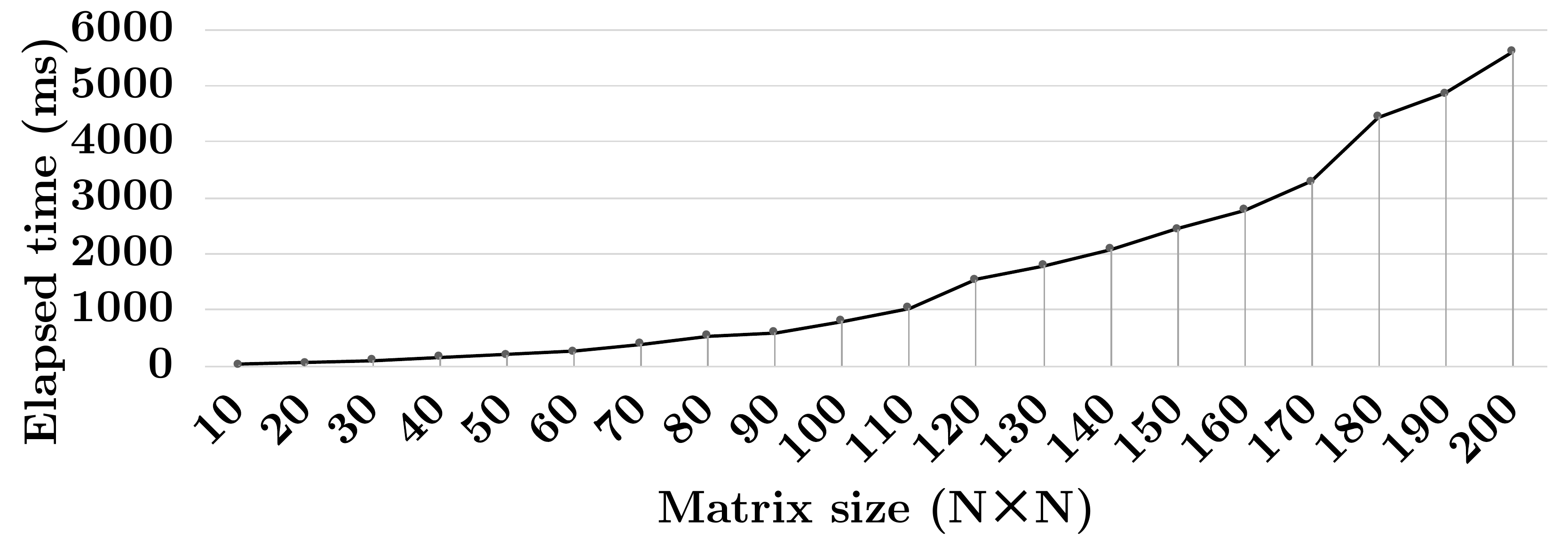}
	\caption{Performance measurements of matrix creation, 
		X: the number of concepts in the taxonomy, 
		Y: the elapsed time}
	\label{fig:evaluation-matrix-creation}
\end{figure}

For each value of $n$ we create 10 distinct random matrices, and subsequently, average their creation times yielding one value per each $n$.
The outcome of this experiment is plotted in \Cref{fig:evaluation-matrix-creation}.
These results confirm that matrix creation scales linearly in the size of matrices, i.e., $n$.

%
\subsection{Effectiveness}

We use the integrity protection taxonomy as our baseline to evaluate the effectiveness of our keyword matching techniques.
As the first step, we remove all the mapped articles on the taxonomy.
Then, to compare the conformity of the automated imports to the manual ones,
we import the same set of articles using each of the keyword matching techniques.

Throughout the experiment, we set the minimal similarity as constant - 
$0.9$ for Dice's coefficient, $1$ for Levenshtein distance, and $-150$ for Fuzzysort.
For fairness, we define no synonyms for the taxonomy concepts.
In practice, users should use synonyms to further boost the mapping.

\begin{figure}[!htb]
	\includegraphics[clip, trim=0.0cm 0.0cm 0.0cm 0.0cm,width=\linewidth]{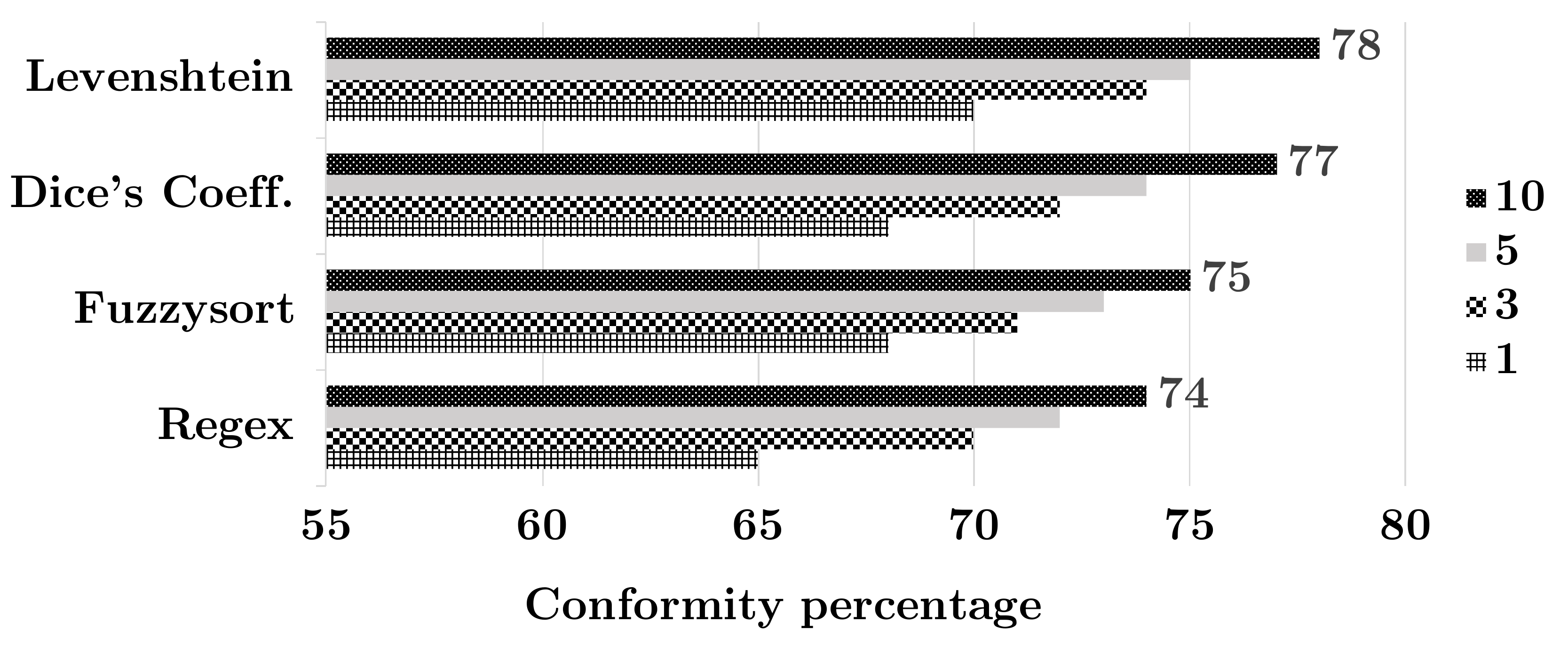}
	\caption{Conformity of the keyword matching techniques to the manually mapped literature with $MOC$ values of 10, 5, 3, 1.}
	\label{fig:evaluation-mapping}
\end{figure}

In our experiments, we define a parameter as \emph{Minimal Occurrence Count} ($MOC$). It dictates how many hits of a concept must appear in a paper  for it to be mapped to the concept. 
As depicted in \Cref{fig:evaluation-mapping}, we experiment the conformity results for four values of $MOC$ 10, 5, 3, and 1.

The results of the Levenshtein distance and the Dice’s coefficient techniques 
have the highest conformity, 78\%, and 77\% respectively.
All of the used string similarity methods seem to perform better than regular expressions (Regex).

\section{Related Work}
\label{sec:related}

Katifori et al. \cite{Katifori07ontologyvisualization} categorize taxonomy visualization techniques based on the visualization concept to \textit{indented list, node-link and tree, zoomable, space-filling, 3D Information landscapes, and Matrix based}. A technique can have functionalities from multiple categories. Most of the existing tools are domain-specific and focus on specific aspects and tasks~\cite{Lohmann2016VisualizingOW}.
In contrast, our platform, besides generic visualization, supports review, analysis, and maintenance tasks. 
Moreover, none of the published techniques in visualization or SLR tools display the complete hierarchy in the matrix, which is crucial for researchers to understand the context of a correlation analysis.

\section{Conclusions}
\label{sec:conclusion}
Our tool chain automates collective taxonomy creation, maintenance and more importantly analysis.
It offers a wide range of tools to aid the identification of research gaps.

We incorporated a set of requirements in the design of our TaaS based on the state of the art and our first-hand experience with developing taxonomies.
Our evaluations indicate that our TaaS is both effective and efficient to be used for developing UML-conforming taxonomies.

As per the future work we plan to support Eclipse Modeling Framework (non-UML relationships) models.
  

\section{Availability}\label{sec:availability}
Our TaaS is freely available at \url{https://www22.in.tum.de/tools/integrity-taxonomy} for the public.
The ease of deployability of our platform makes on-premises solutions another alternative. 
All the source codes are made publicly available on Github at https://github.com/mr-ma/paper-review-go.

\bibliographystyle{ACM-Reference-Format}
\bibliography{bibliography}


\begin{thebibliography}{10}


\ifx \showCODEN    \undefined \def \showCODEN     #1{\unskip}     \fi
\ifx \showDOI      \undefined \def \showDOI       #1{#1}\fi
\ifx \showISBNx    \undefined \def \showISBNx     #1{\unskip}     \fi
\ifx \showISBNxiii \undefined \def \showISBNxiii  #1{\unskip}     \fi
\ifx \showISSN     \undefined \def \showISSN      #1{\unskip}     \fi
\ifx \showLCCN     \undefined \def \showLCCN      #1{\unskip}     \fi
\ifx \shownote     \undefined \def \shownote      #1{#1}          \fi
\ifx \showarticletitle \undefined \def \showarticletitle #1{#1}   \fi
\ifx \showURL      \undefined \def \showURL       {\relax}        \fi
\providecommand\bibfield[2]{#2}
\providecommand\bibinfo[2]{#2}
\providecommand\natexlab[1]{#1}
\providecommand\showeprint[2][]{arXiv:#2}

\bibitem[\protect\citeauthoryear{??}{rel}{[n.d.]}]%
        {reller2016morepublishing}
 \bibinfo{year}{[n.d.]}\natexlab{}.
\newblock \bibinfo{title}{Elsevier publishing – a look at the numbers, and
  more}.
\newblock
\newblock
\urldef\tempurl%
\url{https://www.elsevier.com/connect/elsevier-publishing-a-look-at-the-numbers-and-more}
\showURL{%
\tempurl}
\newblock
\shownote{Accessed: 2018-04-22.}


\bibitem[\protect\citeauthoryear{Ahmadvand, Pretschner, and Kelbert}{Ahmadvand
  et~al\mbox{.}}{2018}]%
        {ahmadvand2018taxonomy}
\bibfield{author}{\bibinfo{person}{Mohsen Ahmadvand},
  \bibinfo{person}{Alexander Pretschner}, {and} \bibinfo{person}{Florian
  Kelbert}.} \bibinfo{year}{2018}\natexlab{}.
\newblock \showarticletitle{A Taxonomy of Software Integrity Protection
  Techniques}.
\newblock \bibinfo{publisher}{Elsevier}.
\newblock
\showISSN{0065-2458}
\urldef\tempurl%
\url{https://doi.org/10.1016/bs.adcom.2017.12.007}
\showDOI{\tempurl}


\bibitem[\protect\citeauthoryear{Gomaa and Fahmy}{Gomaa and Fahmy}{2013}]%
        {gomaa2013survey}
\bibfield{author}{\bibinfo{person}{Wael~H Gomaa} {and} \bibinfo{person}{Aly~A
  Fahmy}.} \bibinfo{year}{2013}\natexlab{}.
\newblock \showarticletitle{A survey of text similarity approaches}.
\newblock \bibinfo{journal}{\emph{International Journal of Computer
  Applications}} \bibinfo{volume}{68}, \bibinfo{number}{13}
  (\bibinfo{year}{2013}).
\newblock


\bibitem[\protect\citeauthoryear{Katifori, Halatsis, Lepouras, Vassilakis, and
  Giannopoulou}{Katifori et~al\mbox{.}}{2007}]%
        {Katifori07ontologyvisualization}
\bibfield{author}{\bibinfo{person}{Akrivi Katifori},
  \bibinfo{person}{Constantin Halatsis}, \bibinfo{person}{George Lepouras},
  \bibinfo{person}{Costas Vassilakis}, {and} \bibinfo{person}{Eugenia
  Giannopoulou}.} \bibinfo{year}{2007}\natexlab{}.
\newblock \bibinfo{title}{Ontology Visualization Methods -- A Survey}.
\newblock
\newblock


\bibitem[\protect\citeauthoryear{Kitchenham and Charters}{Kitchenham and
  Charters}{2007}]%
        {Kitchenham07guidelinesfor}
\bibfield{author}{\bibinfo{person}{B. Kitchenham} {and} \bibinfo{person}{S
  Charters}.} \bibinfo{year}{2007}\natexlab{}.
\newblock \bibinfo{title}{Guidelines for performing Systematic Literature
  Reviews in Software Engineering}.
\newblock
\newblock


\bibitem[\protect\citeauthoryear{Lohmann, Negru, Haag, and Ertl}{Lohmann
  et~al\mbox{.}}{2016}]%
        {Lohmann2016VisualizingOW}
\bibfield{author}{\bibinfo{person}{Steffen Lohmann}, \bibinfo{person}{Stefan
  Negru}, \bibinfo{person}{Florian Haag}, {and} \bibinfo{person}{Thomas Ertl}.}
  \bibinfo{year}{2016}\natexlab{}.
\newblock \showarticletitle{Visualizing ontologies with VOWL}.
\newblock \bibinfo{journal}{\emph{Semantic Web}} \bibinfo{volume}{7},
  \bibinfo{number}{4} (\bibinfo{year}{2016}), \bibinfo{pages}{399--419}.
\newblock


\bibitem[\protect\citeauthoryear{Parreiras, Walter, and Gr\"{o}ner}{Parreiras
  et~al\mbox{.}}{2010}]%
        {parreiras2010visualizing}
\bibfield{author}{\bibinfo{person}{Fernando~Silva Parreiras},
  \bibinfo{person}{Tobias Walter}, {and} \bibinfo{person}{Gerd Gr\"{o}ner}.}
  \bibinfo{year}{2010}\natexlab{}.
\newblock \showarticletitle{Visualizing Ontologies with UML-like Notation}. In
  \bibinfo{booktitle}{\emph{Ontology-Driven Software Engineering}}
  \emph{(\bibinfo{series}{ODiSE'10})}. \bibinfo{publisher}{ACM},
  \bibinfo{address}{New York, NY, USA}, Article \bibinfo{articleno}{4},
  \bibinfo{numpages}{6}~pages.
\newblock
\showISBNx{978-1-4503-0548-8}
\urldef\tempurl%
\url{https://doi.org/10.1145/1937128.1937132}
\showDOI{\tempurl}


\bibitem[\protect\citeauthoryear{Petersen, Feldt, Mujtaba, and
  Mattsson}{Petersen et~al\mbox{.}}{2008}]%
        {petersen2008systematic}
\bibfield{author}{\bibinfo{person}{Kai Petersen}, \bibinfo{person}{Robert
  Feldt}, \bibinfo{person}{Shahid Mujtaba}, {and} \bibinfo{person}{Michael
  Mattsson}.} \bibinfo{year}{2008}\natexlab{}.
\newblock \showarticletitle{Systematic Mapping Studies in Software
  Engineering.}. In \bibinfo{booktitle}{\emph{EASE}}, Vol.~\bibinfo{volume}{8}.
  \bibinfo{pages}{68--77}.
\newblock


\bibitem[\protect\citeauthoryear{Wang and Parsia}{Wang and Parsia}{2006}]%
        {Wang2006CropCirclesTS}
\bibfield{author}{\bibinfo{person}{Taowei~David Wang} {and}
  \bibinfo{person}{Bijan Parsia}.} \bibinfo{year}{2006}\natexlab{}.
\newblock \showarticletitle{CropCircles: Topology Sensitive Visualization of
  OWL Class Hierarchies}. In \bibinfo{booktitle}{\emph{International Semantic
  Web Conference}}.
\newblock


\bibitem[\protect\citeauthoryear{Zowghi and Coulin}{Zowghi and Coulin}{2005}]%
        {zowghi2005requirements}
\bibfield{author}{\bibinfo{person}{Didar Zowghi} {and} \bibinfo{person}{Chad
  Coulin}.} \bibinfo{year}{2005}\natexlab{}.
\newblock \showarticletitle{Requirements elicitation: A survey of techniques,
  approaches, and tools}.
\newblock In \bibinfo{booktitle}{\emph{Engineering and managing software
  requirements}}. \bibinfo{publisher}{Springer}, \bibinfo{pages}{19--46}.
\newblock


\end{thebibliography}
\end{document}